# Towards a high-intensity muon source at CiADS

Han-Jie Cai[1,2], Yuan He[1,2*], Shuhui Liu[1,2], Huan Jia[1,2], Yuanshuai Qin[1], Zhijun Wang[1,2], Fengfeng Wang[1,2], Lixia Zhao[1], Neng Pu[1], Jianwei Niu[1,2,3], Liangwen Chen[1,2], Zhiyu Sun[1,2], Hongwei Zhao[1,2], Wenlong Zhan[1,2,4]

（[1]*Institute of Modern Physics, Chinese Academy of Sciences, Lanzhou 730000*；[2]*School of Nuclear Science and Technology, University of the Chinese Academy of Sciences, Beijing 100049;* [3]*Lanzhou University, Lanzhou 730000;* [4]*Chinese Academy of Sciences, Beijing 100864*）

**Abstract**

The proposal of a high-intensity muon source driven by the CiADS linac, which has the potential to be one of the state-of-the-art facilities, is presented in this paper. We briefly introduce the development progress of the superconducting linac of CiADS. Then the consideration of challenges related to the high-power muon production target is given and the liquid lithium jet muon production target concept is proposed, for the first time. The exploration of the optimal target geometry for surface muon production efficiency and the investigation into the performance of liquid lithium jet target in muon rate are given. Based on the comparison between the liquid lithium jet target and the rotation graphite target, from perspectives of surface muon production efficiency, heat processing ability and target geometry compactness, the advantages of the new target concept are demonstrated and described comprehensively. The technical challenges and the feasibility of the free-surface liquid lithium target are discussed.

## 1. Introduction

The science case of muon sources is extremely rich, covering research fields from fundamental particle physics, nuclear physics and chemistry to condensed matter physics. Moreover, muon sources also have broad applications in elemental analysis and energy research. There are four large-scale facilities providing muons for experiments and user instrumentation around the world. The ISIS muon source at RAL (UK), the SμS at PSI (Swiss), the MUSE facility at J-PARC (Japan) and the CMMS muon source at TRIUMF (Canada) are all active in condensed matter research. In addition, PSI, J-PARC and FNAL (USA) are pursuing broader muon programs in particle physics based on their muon source or dedicated muon beamlines. Several other accelerator facilities such as CSNS (China), RAON (Korea) and SNS (USA) are developing or considering a future muon source [1-3].

Along with the increasing hunger after more intense muon sources to push forward research capabilities and make further discoveries [4,5], substantial efforts have been devoted to development or upgrade of accelerators, targets, beamlines and detectors [6-12]. The High-Intensity Muon Beams (HIMB) project planned at PSI, which aims to obtain an increase by two orders of magnitude in muon rate, is one of the most ambitious projects [13]. The completely unprecedented surface muon intensity of $10^{10}$ μ$^+$/s will allow for completely new experiments with considerable discovery potential and unique sensitivities in particle physics, condensed matter physics and materials science [5].

Under the current beam power limit of 1.4 MW, to achieve the upgrading goals of the muon intensity, the attention of HIMB project is turned to the optimization of existing target stations and beamlines [14,15]. First of all, the target station will be completely rebuilt. The effective interaction length of the new production target will be larger and the geometry will be more optimized to produce more surface muons. More importantly, the muon beamlines will be changed to be based on large-aperture solenoids and dipoles that can capture and transport the surface muons from target with an overall efficiency of around 10 %. The slanted geometry of the target is able to increase the rate of surface muons by up to 50 % [14] and the upgrade in muon beamlines will improve the total efficiency by a factor of 20 [15].

* Corresponding author.
    E-mail address: hey@impcas.ac.cn (Yuan He).

To further push forward the intensity level of muon source, the development of next generation proton drivers is essential. In fact, a 500-MeV superconducting linac designed for Accelerator-Driven System (ADS) project is now under construction in China [16]. The development dates back to 2011 and the prototype front-end linac (CAFe) reached the milestone of a 20-MeV proton beam with an averaged current of 10 mA in early 2021 [17], successfully demonstrating the feasibility of a superconducting linac in Continuous Wave (CW) mode. In the same year, the project named CiADS (Initiative Accelerator Driven sub-critical System) was launched to build a large-scale ADS experimental facility. The superconducting linac is designed to accelerate a proton beam to 500 MeV with an averaged current of 5 mA [18]. As an experimental facility, the sub-critical reactor of CiADS is designed to operate three months per year. Therefore, the proton beam will be available for the experiment and user terminals in an independent hall as shown in Fig. 1, with a considerable percentage of beam time [16].

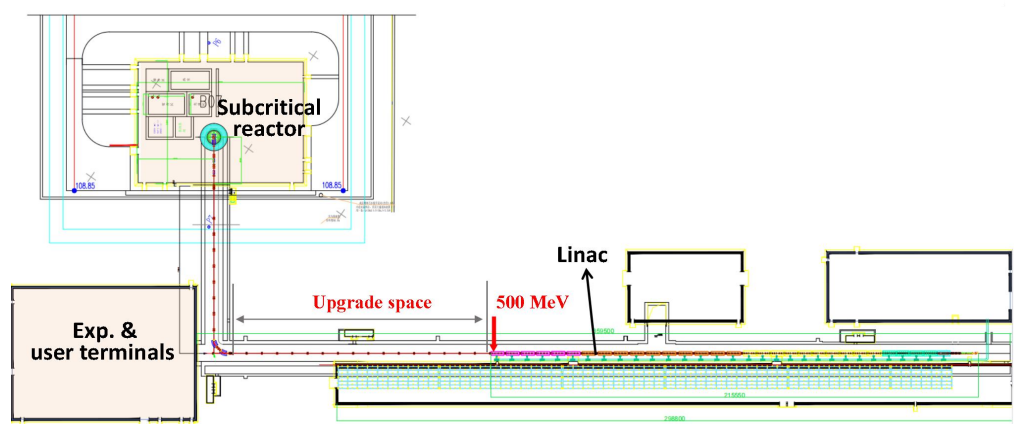

Fig. 1: Layout of the CiADS facilities.

The proposal of a high-intensity muon source driven by the CiADS linac has been considered for several years. From the perspective of beam power, such a muon source has the potential to be one of the state-of-the-art facilities. Technical challenges are certainly exist. The exploration of novel target ideas is essential to meet the challenges posed by the unprecedented beam power. In this paper, a brief introduction to the design, the progress and the upgrade plan of the linac is given in Section 2. This is followed by Section 3, where the consideration of challenges related to the high-power muon production and a new target concept based on free-surface liquid lithium jet proposed for CiADS muon source is presented. Section 4 then describes the exploration of the optimal target geometry for surface muon production and the comparison between lithium jet target and rotation graphite target. In Section 5, discussion and conclusion are given.

## 2. Development progress of the CiADS linac

Fig. 2 presents the schematic diagram of the CiADS linac [18], which mainly consists of a normal conducting front-end, a superconducting acceleration section and several High Energy

\* Corresponding author.
 E-mail address: hey@impcas.ac.cn (Yuan He).

Beam Transport (HEBT) lines. The front-end consists of an Electron Cyclotron Resonance (ECR) ion source, a Low Energy Beam Transport (LEBT) containing a fast chopper for beam pulse structuring and machine protection, a Radio Frequency Quadrupole (RFQ) linac and a Medium Energy Beam Transport (MEBT). The proton beam out of front-end is a CW beam with a current of 5 mA and an energy of 2.1 MeV. After MEBT, the superconducting section takes charge of the acceleration from 2.1 MeV to 500 MeV. Three kinds of Half Wave superconducting Resonators (HWR010, HWR019 and HWR040) and two types of elliptical cavities (Ellip062 and Ellip082) are housed in 32 cryomodules in the superconducting section [19].

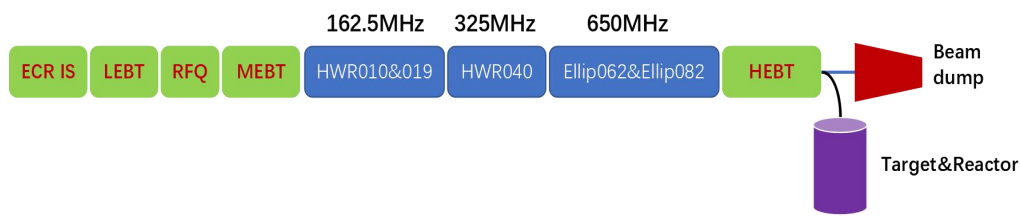

Fig. 2: Schematic diagram of the CiADS linac.

The civil construction of CiADS project is still underway [16] and the front end of the linac had been integrated in December 2022 in a temporary shed, as shown in Fig. 3. The beam commissioning has been carried out successfully with a proton beam out of the RFQ with an energy of 2.18 MeV and a current of 5.2 mA. According to the schedule of the CiADS project [16], the 500 MeV proton beam will achieved by 2025, with a current of 50 μA, and power ramping to 250 kW and 2.5 MW will be realized by 2027 and 2029, respectively. Along with the muon source proposal, the upgrade of the superconducting linac to 600 MeV is also under consideration. In fact, the length of the linac tunnel was designed to allow the upgrade of the linac to an energy of no less than 1.5 GeV at the beginning, as shown in Fig. 1.

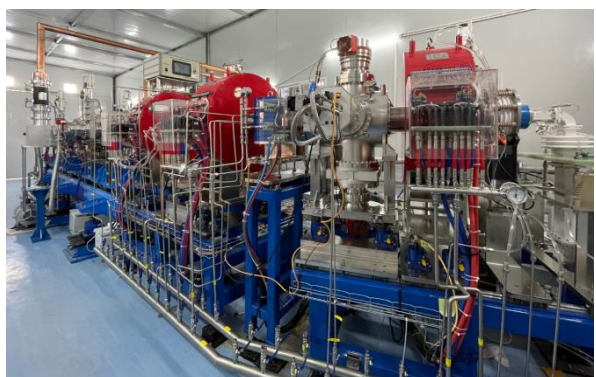

Fig. 3: Photograph of the front-end of CiADS linac in a temporary shed.

## 3. New target concept for the CiADS high-intensity muon source

The production target of a high-intensity muon source is usually challenging owing to high-heat density and harsh irradiation environment [20]. From the target, typical collecting object is the so-called surface muons possessing beneficial properties which can be used by diverse experiments. Originating from the tow-body decay of the positive pion stopped close to the surface of the production target, surface muons escape the target with a momentum ranging from 0 to 29.8 MeV/c. To facilitate the escape of muon, the production target of a muon source is usually thin and direct water cooling by forced-convection on target surface is usually excluded.

Rotation graphite target cooled by thermal radiation is currently the principal candidate for the high-intensity muon sources driven by a proton beam with beam power of several hundreds of kWs or higher, such as SμS, MUSE and the proposed ROAN μSR facility [21-23]. Graphite is chosen for its thermal, mechanical and low-activation properties and its high efficiency in muon production [24,25]. Based on the operation experience of the rotation target at PSI, the main disadvantage of the rotation target is the limited lifetime of the bearings which have to operate without grease due to the harsh irradiation environment. Owing to the degradation of the bearings from heat and radiation, The hub of target E at PSI have to be exchanged by remote-handling manipulators in hot cell after a few months of operation [21]. For the planned HIMB facility, the distance between the target and the capture solenoid is as short as 25 cm for a higher capture efficiency and the tight space constraint certainly will pose restrictions on the optimization of the rotation target system [26].

\* Corresponding author.
　E-mail address: hey@impcas.ac.cn (Yuan He).

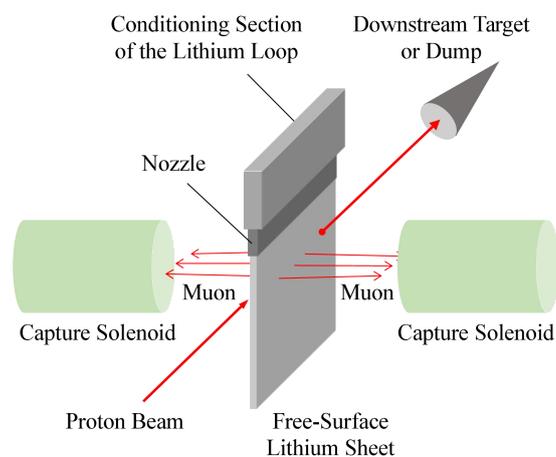

Fig. 4: Schematic diagram of the free-surface liquid lithium target.

Here the novel concept based on a free-surface, sheet-shaped liquid lithium target is proposed. As illustrated in Fig. 4, pressured liquid lithium flows through the conditioning section of a lithium loop and finally forms a sheet-shaped jet out of the narrow nozzle. The proton beam is collimated to hit the lithium jet under a small angle and surface muons produced in lithium escape from either side of the sheet, entering the capture field of the solenoids.

Owing to its properties of low melting point, high saturated vapor pressure, high heat capacity and good compatibility with structural materials, liquid lithium has been widely used as neutron production target [27,28], radionuclide production target[29] and ion beam charge stripper [30]. For the high-intensity muon source driven by CiADS linac, liquid lithium target will be an excellent choice not only due to the properties mentioned above but also because of its low atomic number. The research performed at PSI indicated that the surface muon production efficiency is approximately proportional to $Z^{-2/3}$ with Z being the atomic number [31]. Therefore it is can be expected that liquid jet target will perform well in muon production efficiency. This assumption will be confirmed in Section 4.

A drawback of this target concept is related to the low density of lithium. With a density of 0.515 $g/cm^3$, the effective thickness of lithium target would be several times larger than that of graphite target to keep a same proton beam utilization rate. A longer target tends to lead to a larger emittance which is likely to be detrimental to capture and transmission of muon beam. Therefore, it is necessary to investigate the influence of the target thickness on the muon rate can be obtained.

The simulation studies of muons production and capture processes were performed using the

* Corresponding author.
  E-mail address: hey@impcas.ac.cn (Yuan He).

FLUKA program [32]. With two normal conducting solenoids being placed symmetrically at a distance of 25 cm to the lithium sheet, the investigation into surface muon rates before and after the solenoid are performed for different target thickness in proton beam direction. The capture magnetic field in the horizontal cross section at proton beam height is shown in Fig. 5(a). Both aperture and length of the solenoid are 50 cm. In the simulation studies presented in this section, the lithium slab is set to be hit by proton beam without rotation and the magnetic field produced by the two capture solenoids together is applied. The distribution of the axial magnetic field ($B_r$) along the central axis of capture solenoids is shown in Fig. 5(b). It can be seen that the maximum field is ~ 0.4 Tesla, which is smaller than the limiting magnetic field of a normal conducting solenoid with the current design [26].

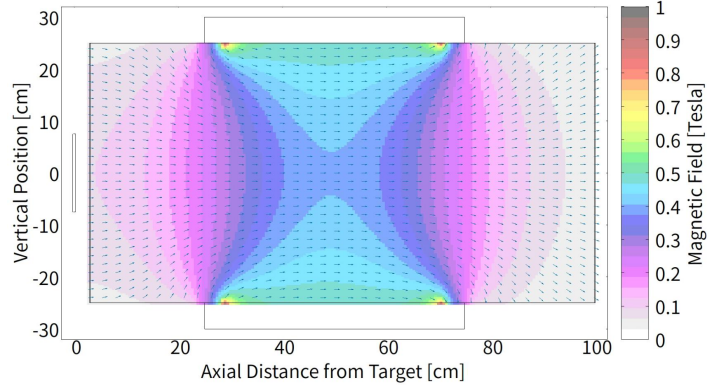

(a)

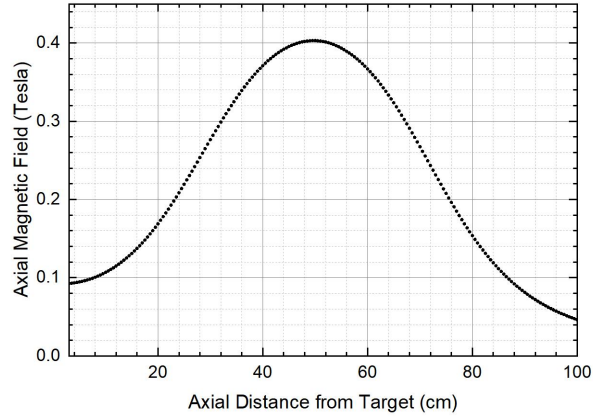

(b)

Fig. 5: (a) Capture magnetic field map in the horizontal cross section at proton beam height. (b) Distribution of axial magnetic field ($B_r$) along the central axis of the capture solenoid.

The $\mu^+$ from target is recorded by a virtual detector placed centrally at a distance of 3 cm from the target, parallel to the proton beam. Another virtual detector is placed at 100 cm from the target, on the downstream side of the solenoid to detect the captured $\mu^+$. The diameters of both detectors are 50 cm, same with the solenoid aperture. A Gaussian proton beam with a $\sigma_x/\sigma_y$ of 1 mm and an energy of 600 MeV is used to impact on the lithium slab target with a fixed width of 10 mm while varying in length. The proton beam current is 5 mA. Positive muons recorded by the two detectors are normalized to the target length to get $NI_{det1}$ and $NI_{det2}$. A filter of $P_{\mu^+} <$ 30 MeV/c where $P_{\mu^+}$ is the momentum of $\mu^+$ is implemented here for both $NI_{det1}$ and $NI_{det2}$.

* Corresponding author.
  E-mail address: hey@impcas.ac.cn (Yuan He).

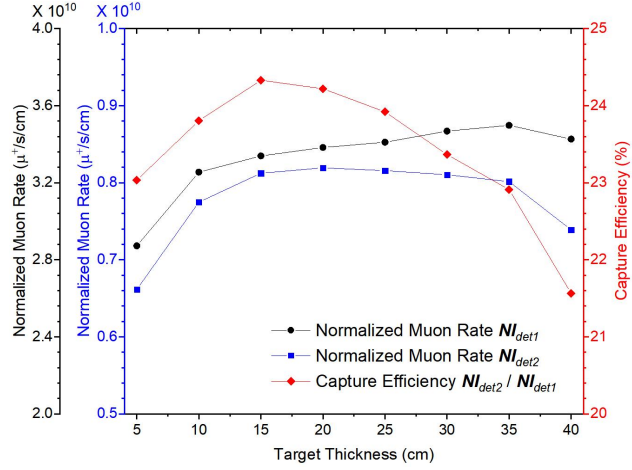

Fig. 6: The normalized $\mu^+$ rates and the capture efficiencies as functions of lithium target thickness. The normalized $\mu^+$ rates $NI_{det1}$ and $NI_{det2}$ are given by the 1-st left scale and the 2-st left one, respectively. The capture efficiency $NI_{det2}/NI_{det1}$ is given by the right scale.

As shown in Fig. 6, the normalized surface muon rate from target $NI_{det1}$ increases to the maximum of $3.5 \times 10^{10}$ μ$^+$/s/cm by a percentage of more than 20 % when the target thickness varies from 5 cm to 35 cm. This should mainly be owed to the increase of the percentage of side-facing muons. The coulomb scattering of proton beam by target also can facilitates the escape of surface muons from target. For a larger target thickness of 40 cm, normalized surface muon rate turns to decrease slightly because of the increase of emittance. For comparison, the turning point of $NI_{det2}$ appears at the target thickness of 20 cm where the muon rate is $0.82 \times 10^{10}$ μ$^+$/s/cm, earlier than that of $NI_{det1}$. The capture efficiency defined by $NI_{det2} / NI_{det1}$ increases to the maximum of 24.3 % at the target thickness of 15 cm. More generally, both the captured $\mu^+$ rate and the capture efficiency vary by less than 7 % in a wide thickness range from 10 cm to 35 cm. Therefore, it is reasonable to believe that the low density of lithium will not result in significant disadvantage in capture efficiency for a large-aperture solenoid-based beamline which will be used by next-generation muon sources like HIMB [26].

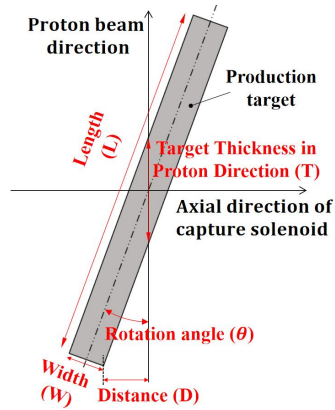

Fig. 7: Schematic diagram of the beam-target geometry with key dimensions denoted and marked in red.

## 4. Performance of liquid lithium jet target in muon rate

It has been demonstrated that a gain in surface muon rate can be achieved by implementing a small rotation angle on slab target [26,31]. Fig. 7 illustrates the beam-target geometry for slanted slab target with key dimensions denoted and marked in red. There are equations listed as follow for the dimensions:

$$W \leq L * \tan\theta - 2 * D/\sin\theta \quad (1)$$
$$L = T * \cos\theta + 2 * D/\sin\theta \quad (2)$$
$$T = W/\sin\theta. \quad (3)$$

\* Corresponding author.
  E-mail address: hey@impcas.ac.cn (Yuan He).

With a jet length L of 25 cm and a distance D of 5 mm to guarantee the space for proton beam on the liquid lithium jet target, the maximum target width and the corresponding target thickness in proton beam direction as functions of rotation angle θ are shown in Fig. 8.

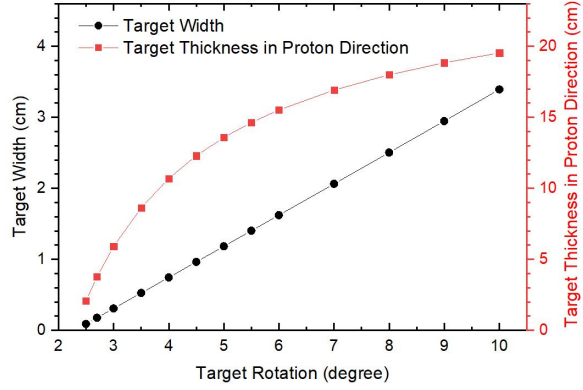

Fig. 8: The maximum target width (left scale) and the maximum target thickness in proton beam direction (right scale) as functions of target rotation angle.

It is easy to understand that the increase of the effective thickness tends to be less sharp for a larger rotation angle. As a result, the gain in muon rate from the increase of rotation angle will be less significant. With the same configurations as that implemented in Section 3 for proton beam, capture solenoid and detectors, the investigations into μ+ rate recorded by the detector downstream the solenoid were undertaken for the liquid lithium jet targets with different rotation angles.

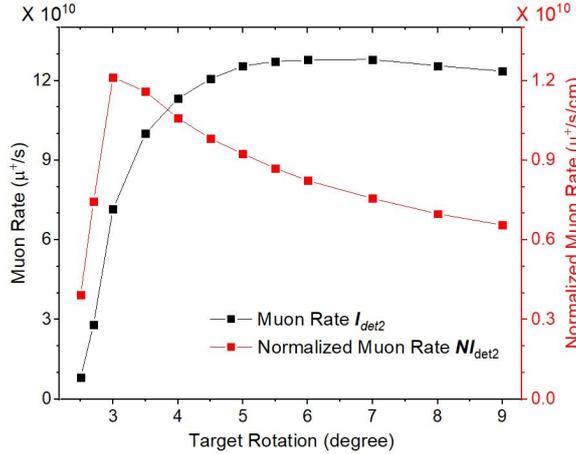

Fig. 9: The μ+ rates at the second detector as functions of target rotation angle. That given by right scale is normalized to the target thickness in proton beam direction.

As shown in Fig. 9, the μ+ rate $I_{det2}$ slowly increases to the maximum of $12.8 \times 10^{10}$ μ+/s at the angle of 6 degree and tends to decrease when rotation angle increase further. In addition to thickness, the width which determines both pion stopping rate and muon escaping efficiency is also crucial to μ+ rate. A small target width will result in a small pion stop rate while a large one is adverse for muon escaping. From the perspective of μ$^+$ rate per unit target thickness, the optimal rotation angle is 3 degree, where the effective thickness is 5.9 cm and the normalized μ$^+$ rate is $1.2 \times 10^{10}$ μ$^+$/s/cm. The decrease after the rotation angle of 3 degree indicates that the detrimental effect of a large target width is dominating the μ$^+$ production efficiency.

* Corresponding author.
  E-mail address: hey@impcas.ac.cn (Yuan He).

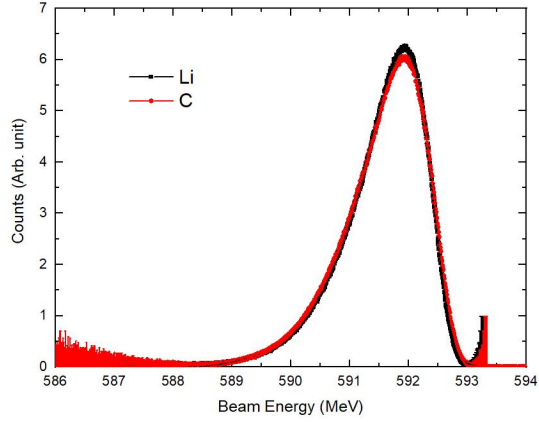

Fig. 10: Energy distributions of the transmitted proton beams after the 7.8-cm lithium and the 2-cm graphite.

For the comparison of the liquid lithium target to the rotation graphite target, a rotation angle of 3.5 degree and an effective thickness of 7.8 cm are set for lithium target while the configuration with a 10-degree rotation and a 2-cm effective length are used for graphite slab, which is also the baseline configurations of the HIMB graphite target. As shown in Fig. 10, the energy distributions of proton beams are almost the same after penetrating through the two targets. Therefore, it is reasonable to assume that the downstream beam losses caused by energy loss and multiple scattering in target will be very close.

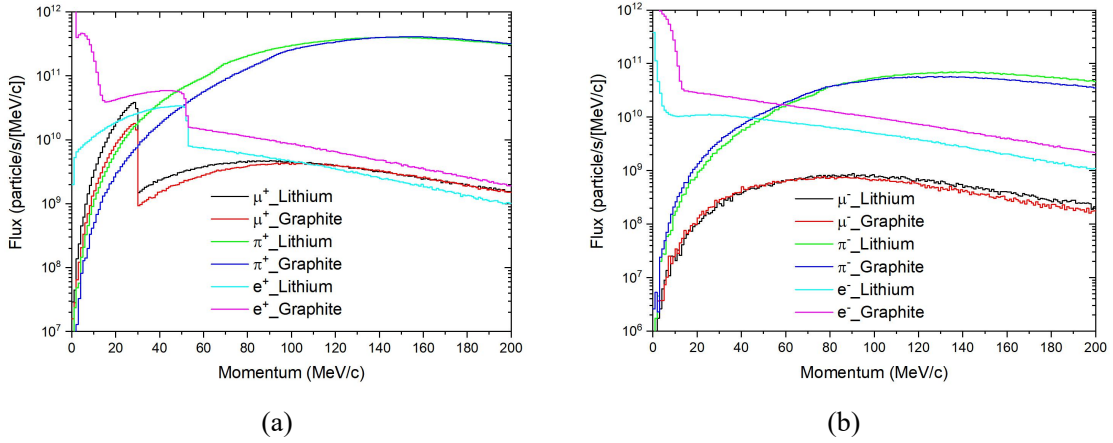

(a)                     (b)

Fig. 11: Momentum spectra of positive (a) and negative particles (b) recorded by the detector upstream the capture solenoids.

Fig. 11 presents the momentum spectra of positive and negative particles recorded by the detector upstream the capture solenoids for both lithium and graphite target. It is can be seen that the yields of $\mu^+$ and $\pi^+$ from lithium is larger than that from graphite, especially in the momentum range from 0 to 100 MeV/c. For $\mu^-$ and $\pi^-$, the yields from two targets are very close. Another advantage of lithium target is that the substantial backgrounds from positrons and electrons are about one order of magnitude lower, which will make the background separation

* Corresponding author.
  E-mail address: hey@impcas.ac.cn (Yuan He).

requirements less challenging.

The surface muon rates from the 7.8-cm lithium target are $3.82 \times 10^{10}$ μ$^+$/s and $9.27 \times 10^{10}$ μ$^+$/s at detector 1 and detector 2, respectively. Both are 2.1 times that from the 2-cm graphite target, which means that the capture efficiencies for two target are almost the same. From the perspective of the distributions in phase space and momentum-polarization space, as shown in Fig. 12, all of the distributions for both targets are very close except for horizontal position at detector 1. The standard deviation $\sigma_x$ for lithium is observably larger than that for graphite due to the difference in target thickness.

Basing on the fact that target thickness makes little difference in the capture efficiency of surface muons until increasing to be comparable with solenoid aperture, as already demonstrated in Section 3, and the high similarity in the distributions in phase spaces and momentum-polarization space at detector 2, we now can come to a conclusion that the surface muon rate can be doubled, without loss in downstream transmission efficiency, by using liquid lithium target instead of rotation graphite target. As demonstrated in Fig. 12, for both targets, the polarization of the muon beam from lithium target increase to 0.95 after the capture solenoid by the reduction in both vertical and horizontal divergence.

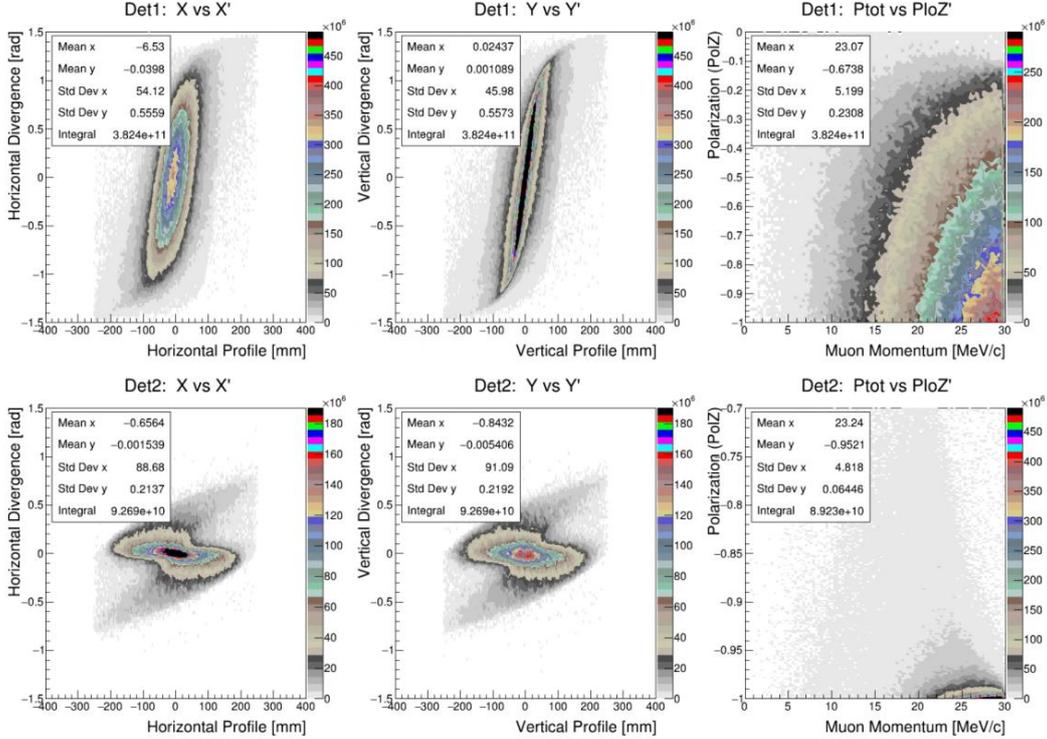

(a)

* Corresponding author.
E-mail address: hey@impcas.ac.cn (Yuan He).

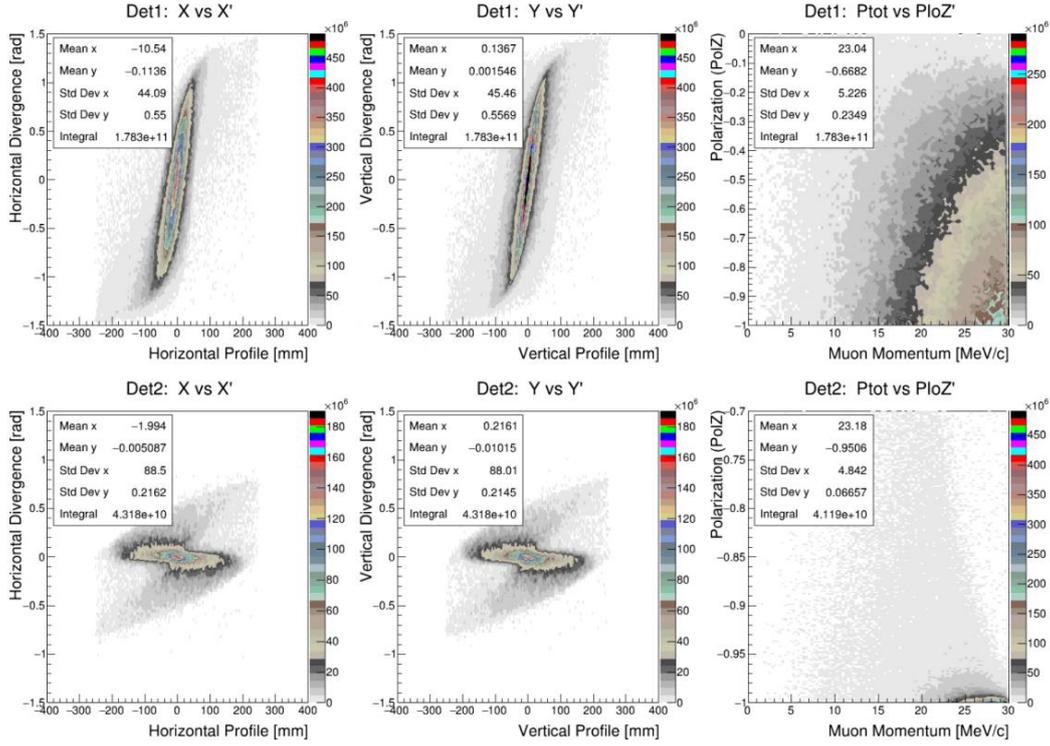

(b)

Fig. 12: (a) Distributions of surface muons in horizontal phase space (left), vertical phase space (middle) and momentum-polarization space (right) from the 7-cm lithium target. Top three for produced surface muons recorded by detector 1 and bottom ones for captured surface muons recorded by detector 2. (b) Same as (a) except that the target is the 2-cm graphite.

## 5. Discussion and conclusion

Unlike rotation graphite target, liquid lithium target is free from the restrictions posed by space limitations and manufacturing of the graphite wheel. Therefore, it is realistic to increase the effective thickness further for a higher proton beam utilization rate and thus a higher muon beam intensity. With the 5-mA proton beam of CiADS linac, an optimized liquid lithium target and a solenoid-based muon beamline with a transmission efficiency of 40 %, a surface muon rate of more than $5 \times 10^{10}$ μ$^+$/s can be delivered to experiments.

With a rotation of 3.5 degree for the liquid sheet and a $\sigma_x/\sigma_y$ of 1 mm for the 5-mA proton beam, the maximum heat density integrated in jet direction is 21.75 kW/cm$^2$. With a jet velocity of 4 m/s, the maximum local temperature rise of the liquid lithium target is estimated to be less than 30 ℃. This temperature rise is quite small and excessive vaporization of the liquid lithium can be avoided.

The main challenge consist in maintaining the stability of the free-surface liquid lithium sheet. Fortunately, numerous R&D efforts have been devoted to investigate the feasibility of producing free-surface liquid lithium films or sheets for the applications in charge stripping [33], radionuclide production [29] and Inertial Fusion Energy (IFE) reactor chamber first walls protection [34]. At Michigan State University (MSU), a 10~20 μm thick liquid lithium jet flowing at > 50 m/s was created and confirmed stable when bombarded by various heavy ion beams [30]. To demonstrate the feasibility of a windowless lithium target for the Rare Isotope Accelerator (RIA) project, a liquid lithium jet with a cross section of 5 mm × 10 mm and a velocity of being varied up to 6 m/s was produced at Argonne National Laboratory (ANL) and thermal loads of up to 20 kW were applied on the jet by 1-MeV electron beams. It was demonstrated that the free-surface liquid lithium target flowing at a velocity of 1.8 m/s can operate stably without disruption or excess vaporization [35]. To investigate the stability of the high-speed liquid curtain shielding concept, which was proposed in the High Yield Lithium-Injection Fusion Energy (HYLIFE-II) design to protect the first walls from damaging radiation, a series of experiments using water as the simulant have been carried out. The free-surface fluctuations were quantified to be less than 5 % in a wide range of distance from nozzle exit and flow dynamic parameters [36,37].

\* Corresponding author.
  E-mail address: hey@impcas.ac.cn (Yuan He).

Thanks to the research efforts mentioned above, the feasibility of the free-surface liquid lithium target concept, to a certain extent, has been demonstrated given that the Re-We parameter space (Reynolds number and Weber number) of the demanded liquid jet with a width of several millimeters and a velocity of several m/s is well covered by these experiments. An additional doubt is the influence on the lithium jet from the fringing field produced by the capture solenoids. The advantage of the lithium target is a larger space for the mirror plates in front of the capture solenoid to reduce the magnetic field seen by target. In the worst situation where a stable liquid jet is unachievable due to the existence of a partial canceled fringing field, the polarity of two capture solenoids can be set to be inverse and thus their magnetic fields at target shall cancel if the symmetry is well fulfilled.

In summary, the feasibility of the lithium jet target concept, which possesses advantages in surface muon production efficiency, heat processing ability and target geometry compactness, will be further investigated along with the conceptual design of the CiADS muon source. The traditional rotation graphite target is certainly still a candidate solution given that the design goals including muon rate, target number and proton beam consumption rate are not definite yet. However, the integration of the CiADS superconducting CW linac and the free-surface liquid lithium target solution certainly would significantly put forward the intensity frontier of muon sources.

**Acknowledgments**

Fruitful discussions with Professor Jingyu Tang are gratefully acknowledged. This work is supported by the Large Research Infrastructures of 12th Five-Year Plan: China initiative Accelerator Driven System.

\* Corresponding author.
E-mail address: hey@impcas.ac.cn (Yuan He).

\* Corresponding author.
    E-mail address: hey@impcas.ac.cn (Yuan He).

* Corresponding author.
 E-mail address: hey@impcas.ac.cn (Yuan He).